\documentclass[letterpaper,twocolumn,english,aps,prl,groupedaddress,superscriptaddress,showpacs,amsfonts] {revtex4}

\usepackage{epsfig}
\usepackage{graphicx}
\usepackage{dcolumn}
\usepackage{bm}

\usepackage{color}
\usepackage{amsmath}
\usepackage{amssymb}

\renewcommand{\vec}[1]{\boldsymbol{#1}}

\DeclareMathAlphabet{\mathpzc}{OT1}{pzc}{m}{it}


\newcommand{\beq}{\begin{equation}}
\newcommand{\eeq}{\end{equation}}
\newcommand{\bea}{\begin{eqnarray}}
\newcommand{\eea}{\end{eqnarray}}
\newcommand{\lhyf}{LiHo$_x$Y$_{1-x}$F$_4$ }

\begin{document}

\title{Spin Glass Transition at Nonzero Temperature in a Disordered Dipolar Ising System: \\
The Case of LiHo$_x$Y$_{1-x}$F$_4$}


\author{Ka-Ming Tam}

\affiliation {Department of Physics and Astronomy, University of Waterloo, Waterloo, ON, N2L 3G1, Canada}

\author{Michel J. P. Gingras}

\affiliation {Department of Physics and Astronomy, University of Waterloo, Waterloo, ON, N2L 3G1, Canada}

\affiliation {Canadian Institute for Advanced Research, 180 Dundas Street West, Suite 1400, Toronto, ON, M5G 1Z8, Canada}

\date{\today}

\begin{abstract}

The physics of the
spin glass (SG) state, with magnetic moments (spins) frozen in random orientations, 
is one of the most intriguing problems in condensed matter 
physics.  While most theoretical studies have focused on the Edwards-Anderson model
 of Ising spins with only discrete `up/down' directions, such 
Ising systems are experimentally rare. LiHo$_{x}$Y$_{1-x}$F$_{4}$, 
where the Ho$^{3+}$ moments 
are well described by Ising spins, is an almost ideal
Ising SG 
material. In LiHo$_{x}$Y$_{1-x}$F$_{4}$, the Ho$^{3+}$ moments 
interact predominantly via the inherently frustrated 
magnetostatic dipole-dipole interactions. 
The random frustration causing the SG behavior 
originates from the random 
substitution of dipole-coupled Ho$^{3+}$ by non-magnetic Y$^{3+}$. 
In this paper, we provide compelling evidence from extensive computer 
simulations that a 
SG transition at nonzero temperature occurs 
in a realistic microscopic model of LiHo$_{x}$Y$_{1-x}$F$_{4}$, hence resolving 
the long-standing, and still
ongoing, controversy about the existence of a 
SG transition in disordered dipolar Ising systems.

\end{abstract}

\pacs{75.40.Cx,75.40.Mg,75.50.Lk}

\maketitle

The early 1970s discovery of materials failing to develop conventional long-range 
magnetic order down to zero temperature, but displaying a cusp in the 
magnetic susceptibility signaling a transition to a state of randomly frozen spins~\cite{Canella}, 
spurred thirty years of immense theoretical effort aimed at 
understanding these fascinating spin glass (SG) systems~\cite{Binder,Mydosh,Young-book}. 
 In that context, the Edwards-Anderson (EA) model of spins interacting via exchange 
interactions $J_{ij}$, which can be either ferromagnetic or antiferromagnetic 
and chosen from a frozen (quenched) probability distribution function,
$P(J_{ij})$, has been the subject of innumerable theoretical studies. 
Because of the added simplicity of considering Ising spins with only two 
discrete `up/down' orientations, the EA Ising model has attracted particular 
attention. 
However, because Ising magnetic materials are quite 
uncommon, most experimental studies have targeted systems where the moments are 
described instead by isotropic (Heisenberg) spins that can point in 
any direction~\cite{Binder,Mydosh,Young-book}. 
For an Ising description to apply, the single-ion anisotropy energy scale 
must be much larger than the spin-spin interactions. 
This 
often 
occurs in materials where the magnetic species consist of $4\rm f$ 
rare-earth elements such as Tb, Ho or Dy. From that 
perspective, the  \lhyf insulator has long proven to be a remarkable
 material to explore collective magnetic phenomena 
\cite{Reich,Brooke1,Ghosh,Silevitch,Brooke2,Wu}, 
including  SG behavior within an Ising setting \cite{LHYF-Ising,Chakraborty}.

Because of the compactness of the spin-carrying $4\rm f$ orbitals, 
magnetic exchange and superexchange 
between Ho$^{3+}$ ions is small in  \lhyf
and magnetostatic dipolar interactions are the predominant Ho$^{3+}-$Ho$^{3+}$ couplings. 
Also, since the single-ion 
crystal field anisotropy is 
 large compared to the magnetic interactions, 
the Ho$^{3+}$ magnetic moments can be mapped onto effective Ising 
spins that can only point parallel or antiparallel to the 
$c$-axis of the tetragonal crystalline structure 
of LiHo$_{x}$Y$_{1-x}$F$_{4}$  \cite{Chakraborty}.
 Ignoring the 
small nearest-neighbor exchange, which does not qualitatively affect the physics at small $x$, 
\lhyf can thus be described by a 
model of classical Ising spins coupled by long-range dipolar interactions whose Hamiltonian is:

\begin{eqnarray}
H=\frac{D}{2} \sum_{i \neq j} \epsilon_{i} 
\epsilon_{j}\frac{r_{ij}^{2} - 3 z_{ij}^{2}}{r_{ij}^{5}} \sigma_{i} \sigma_{j} \; .
\label{Hamiltonian}
\end{eqnarray}

Here $D>0$ is the scale of the dipolar interactions and $r_{ij}=|\vec{r}_{ij}|$,
 where $\vec{r}_{ij}=\vec{r}_{i}-\vec{r}_{j}$, 
with $\vec{r}_{i}$ and $\vec{r}_{j}$ the positions of Ho$^{3+}$ ions $i$ and $j$.
$z_{ij}\equiv \vec {r}_{ij}\cdot \hat z$ with $\hat z$ parallel to the $c$-axis. 
$\epsilon_{i}=1$ if $\vec{r}_{i}$ is occupied by a magnetic Ho$^{3+}$ ion
and $\epsilon_{i}=0$ otherwise. The Ising variable 
$\sigma_{i} = \pm 1$ for a Ho$^{3+}$ moment pointing along $\pm \hat{z}$, respectively. 
Depending on the relative positions of two interacting
moments, the pairwise $J_{ij} \equiv D(r_{ij}^{2}-3z_{ij}^{2})/r_{ij}^{5}$ interaction 
can be either negative (ferromagnetic) or positive 
(antiferromagnetic). Despite the resulting
 geometrical frustration, pure LiHoF$_{4}$
 exhibits long-range dipolar ferromagnetic order below a critical 
temperature of $T_{c} \approx 1.53$ K \cite{Cooke,Chakraborty,Xu,Biltmo,LHYF-Ising}.
 As Ho$^{3+}$ is progressively substituted by non-magnetic Y$^{3+}$, 
$T_{c}$ decreases, while random frustration concomitantly builds up until, 
for $x_{c} \approx 25\%$, dipolar Ising ferromagnetism disappears \cite{Reich,Biltmo}.

It had long been thought that a dipolar Ising SG state exists in
\lhyf for $x=16.5 \%$ \cite{Reich} while for $x=4.5 \%$,
a mysterious antiglass state occurs \cite{Ghosh,Reich}, 
perhaps due to quantum effects \cite{Ghosh}. It has however recently been 
suggested, 
on the basis of an analysis of the nonlinear magnetic susceptibility, 
that a SG phase might not actually be realized in
\lhyf for $x=16.5 \%$ \cite{Jonsson1}. Even more recent work disputes 
this claim \cite{Ancona-Torres}, not without having
 generated a debate \cite{Jonsson2,Quilliam2}.
 To compound this controversy, all recent numerical 
studies of diluted dipolar Ising models fail 
to find a SG transition at nonzero temperature \cite{Biltmo,Snider}. 
This is in sharp contrast with the long-standing theoretical 
expectations that a transition 
should occur in this system, just as it 
does in the three-dimensional 
(3D) nearest-neighbor EA model~\cite{Ballesteroes,Katzgraber,Hasenbusch}, 
and down asymptotically to $x=0^{+}$ \cite{Stephen}.  
The field is thus faced with a multifaceted conundrum: is there a SG phase in 
diluted dipolar Ising materials  such as 
LiHo$_{x}$Y$_{1-x}$F$_{4}$ \cite{Reich,Jonsson1,Jonsson2,Ancona-Torres,Quilliam2}? 
If not, is the SG phase in \lhyf
destroyed by subsidiary interactions 
responsible for quantum mechanical effects 
that may induce an exotic (e.g. antiglass) 
quantum disordered  state \cite{Ghosh}? 
Or, is the expectation~\cite{Stephen,BM} that 
 random classical dipolar Ising systems ought to exhibit a 
SG transition, just as it does in 3D Ising EA model~\cite{Ballesteroes,Katzgraber,Hasenbusch}, 
simply wrong?
 These are important questions that pertain to our global understanding of 
randomly frustrated systems beyond the celebrated EA model.
 Here, we bring new light on these questions
by investigating model (1) via extensive computer simulations.

We used Monte Carlo simulations to study 
 Eq. \ref{Hamiltonian} 
for a lattice model of 
LiHo$_{x}$Y$_{1-x}$F$_{4}$. 
We considered a tetragonal unit ($C^{6}_{4h}(I4_{1}/a)$ space group) 
with lattice parameters $a=b=5.175$ \AA, $c=10.75$ 
\AA, and with four Ho$^{3+}$ ions per unit cell 
located at $(0,0,c/2)$, $(0,a/2,3c/4)$, $(a/2,a/2,0)$ and $(a/2,0,c/4)$. 
The dipolar coupling $D/a^3$ was set to 0.214 K ~\cite{Biltmo}.
System sizes $La \times La \times Lc$ with $L=6$, $8$, $10$
and an average number $N$ of spins $N=4xL^{3}$ spins 
were investigated via finite-size scaling analysis.
The dipolar lattice sum in (1) was 
performed by summing an infinite 
array of image spins via the Ewald
 method without a demagnetization term \cite{Wang}.

Single spin-flip Monte Carlo simulations using the standard Metropolis 
algorithm is implemented within a parallel thermal tempering scheme \cite{PT1,PT2} 
which has been shown to be highly efficient in speeding up equilibration in glassy systems. 
$N_{T}$ replicas at different temperatures 
were simulated in parallel with consecutive temperatures scaled by a factor $\alpha$. The temperatures
explored for each replica is $T^{(n)} = T_{\rm min}\alpha^{n}$ where $T_{\rm min}$
was the lowest temperature considered and $n \in [0,N_{T}-1]$, thus the highest temperature $T_{\rm max} =
T_{\rm min}\alpha^{N_{T}-1}$ and $\alpha = \root N_{T}-1 \of {T_{\rm max}/T_{\rm min}}$. 
The acceptance ratio for parallel tempering swapping is 
maintained above $50\%$. At least $2\times10^{6}$ Monte Carlo steps ($N_{\rm MCS}$) per 
spin were performed and the last $10^{6}$ of them were used for collecting statistics. 
More than one thousand realizations of disorder 
($N_{\rm sample}$) were considered to perform the disorder average. Table \ref{para_table} 
lists the parameters used in the Monte Carlo 
simulations.

\begin{table}[htbp]
\centering
\begin{tabular}{c c c c c c c c c}
\hline
$x$ & $L$ & $T_{\rm min}$ & $T_{\rm max}$ & $N_{T}$ & $N_{\rm MCS}$ & $N_{\rm samples}$ \\ \hline
$6.25\%$ & 6 & $0.032$K & $0.2$K & $16$ & $2\times10^6$ & 4731 \\
$6.25\%$ & 8 & $0.032$K & $0.2$K & $20$ & $3\times10^6$ & 4057 \\
$6.25\%$ & 10 & $0.032$K & $0.2$K & $24$ & $5\times10^6$ & 2226 \\

$12.5\%$ & 6 & $0.06$K & $0.3$K & $16$ & $2\times10^6$ & 2003 \\
$12.5\%$ & 8 & $0.06$K & $0.3$K & $18$ & $2\times10^6$ & 1822 \\
$12.5\%$ & 10 & $0.06$K & $0.3$K & $24$ & $3\times10^6$ & 1633 \\ \hline 

\end{tabular}
\caption{Parameters of the Monte Carlo simulations.}
\label{para_table}
\end{table}

One way to monitor the  freezing into 
a SG state is to calculate the overlap $q(\vec{k})$ 
of two replicas with the same random 
realization of site occupancy, 
with $q(\vec{k})\equiv\frac{1}{N}\sum_{i=1,2,...,N} \sigma^{(1)}_{i} 
\sigma^{(2)}_{i} \exp(i\vec{k}\cdot\vec{r_{i}})$ and 
where $\sigma^{(1)}_{i}$ and $\sigma^{(2)}_{i}$ are the spins of the two replicas.
 A standard procedure to expose a putative SG
phase transition is to consider the dimensionless (scale-invariant at the critical point) 
Binder ratio  \cite{Ballesteroes,Katzgraber,Hasenbusch,Bhatt},
 $ B = \frac{1}{2}\left (3-\frac{[\langle q^4(0) \rangle ]}{[\langle q^2(0) \rangle ]^2}\right )$,
where $\langle \ldots \rangle$ and $[\ldots ]$ denote thermal average and average over the
$N_{\rm samples}$  realizations of random dilution, respectively.


\begin{figure}[bth]
\centerline{
\includegraphics*[height=0.16\textheight,width=0.24\textwidth, viewport=0 0 750 525,clip]{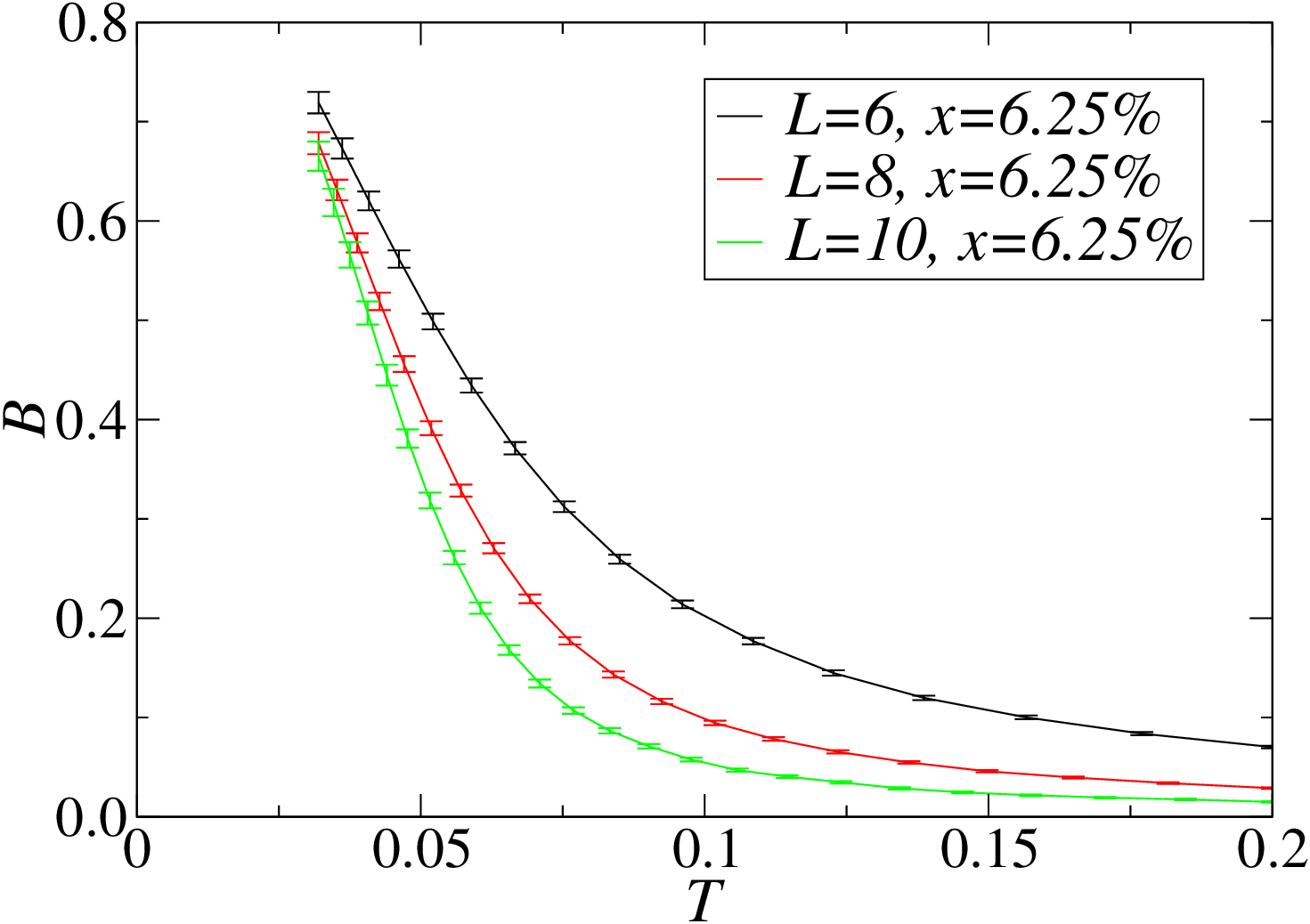}
\includegraphics*[height=0.16\textheight,width=0.24\textwidth, viewport=0 0 750 525,clip]{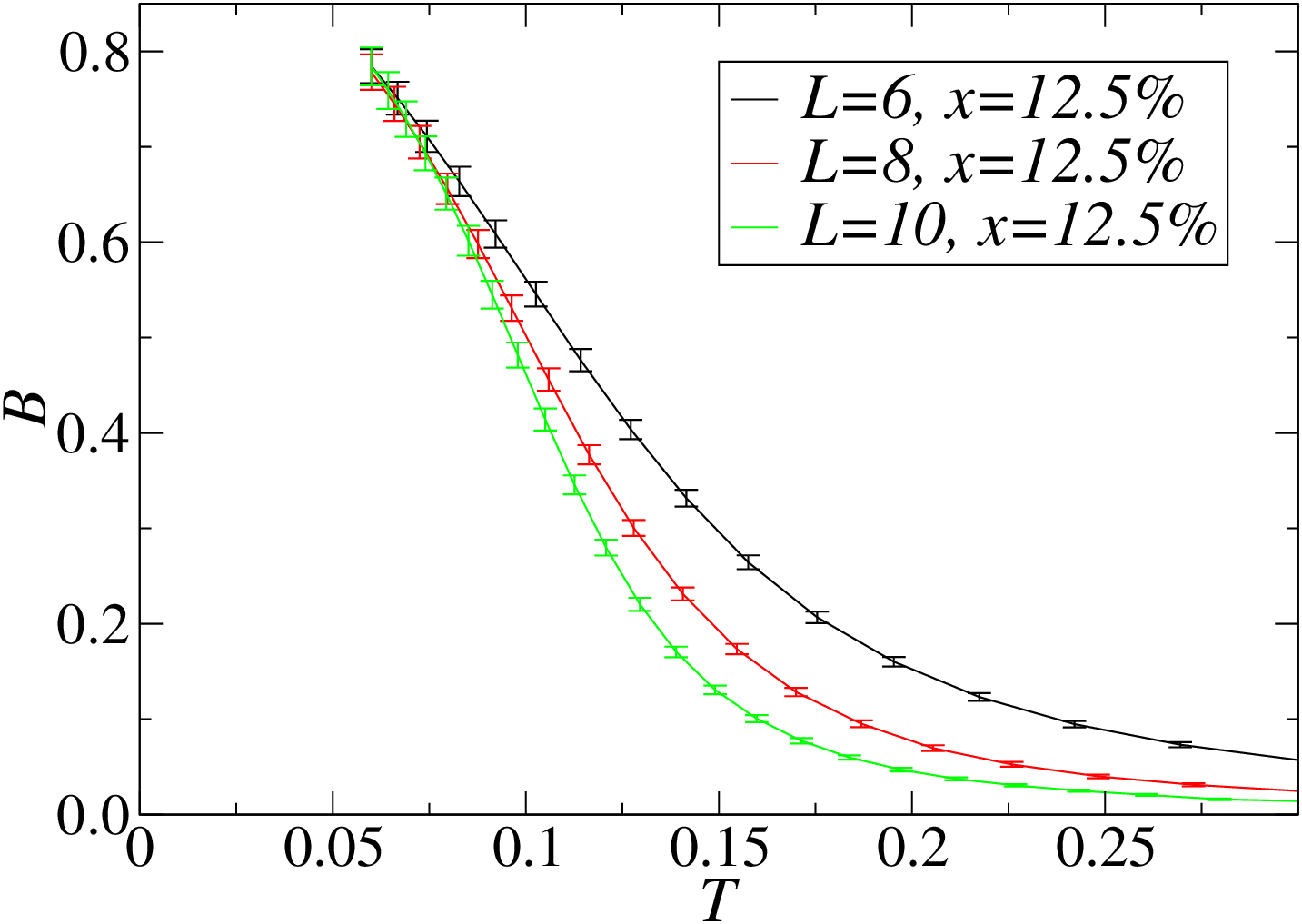} }
\caption{Binder ratios as a function of temperature for $x=6.25\%$(left) and $x=12.5\%$(right), 
where $B$ is dimensionless and $T$ is in K units.}
\label{Binder}
\end{figure}

Figure \ref{Binder} shows $B$ vs temperature $T$ for $x=6.25\%$ and $12.5\%$.  
While $B$ for different system sizes appear to eventually merge below 
a certain temperature, no clear crossing supporting a phase transition can be identified. 
Similar results were recently obtained \cite{Biltmo}, suggesting that no 
finite-temperature SG transition occurs in model (\ref{Hamiltonian}). 
That said, unambiguous $B$ crossings cannot be resolved in many Monte 
Carlo simulations, even for the 3D EA Ising model where a SG transition 
is believed to occur, though, in principle, crossing in $B$ should 
be resolved when the system size is sufficiently large 
\cite{Ballesteroes,Katzgraber,Hasenbusch,Bhatt}. 
Hence, it is perhaps premature to conclude on the basis of results 
such as in Fig. \ref{Binder} that a SG transition does not occur in model (\ref{Hamiltonian}).
Interestingly,  Monte Carlo studies of the 3D EA 
Ising model have found that the SG correlation length, $\xi_{L}/L$, 
is a more suitable 
scale-invariant parameter to expose a possible 
finite-temperature spin freezing transition \cite{Ballesteroes,Katzgraber,Hasenbusch}. 
If a transition occurs, $\xi_{L}/L$ vs  
temperature for different $L$ should cross at $T_{\rm sg}$. 
$\xi_L$ is expected to behave asymptotically for finite $L$ as
$\xi_{L}/L = F[(T-T_{\rm sg})L^{1/\nu}]$,
where $F$ is a universal scaling function. 
The correlation length $\xi_{L}$ above the freezing temperature 
can be approximately determined from the Fourier transform 
of the SG susceptibility, $\chi_{\rm sg}(\vec{k}) \equiv N[q^{2}(\vec{k})]$. 
Assuming that $\chi_{\rm  sg}(\vec{k})$ follows an 
Ornstein-Zernike form above the SG 
transition temperature $T_{\rm sg}$ \cite{Ballesteroes,Katzgraber,Hasenbusch}, 
$\chi_{\rm  sg}(\vec{k}) \propto 1/(\xi^{-2}_{L}+|\vec{k}|^2)$, $\xi_{L}/L$ 
can be determined via 
$\xi_{L} = (\chi_{\rm  sg}({0})/\chi_{\rm sg}(\vec{k})-1)^{1/2}/|\vec{k}|$, 
with $\vec{k}$, chosen as the smallest wave vector for 
the finite-size system, given by $\vec{k}=2\pi \hat {z}/(cL)$.
A suitable form for periodic boundary conditions is
$\xi_{L} = (\chi_{\rm  sg}({0})/\chi_{\rm sg}(\vec{k})-1)^{1/2}/(2 \sin(|\vec{k}|/2))$, 
which we use in the following calculations.

\begin{figure}[bth]
\includegraphics*[height=0.23\textheight,width=0.39\textwidth, viewport=0 0 750 525,clip]{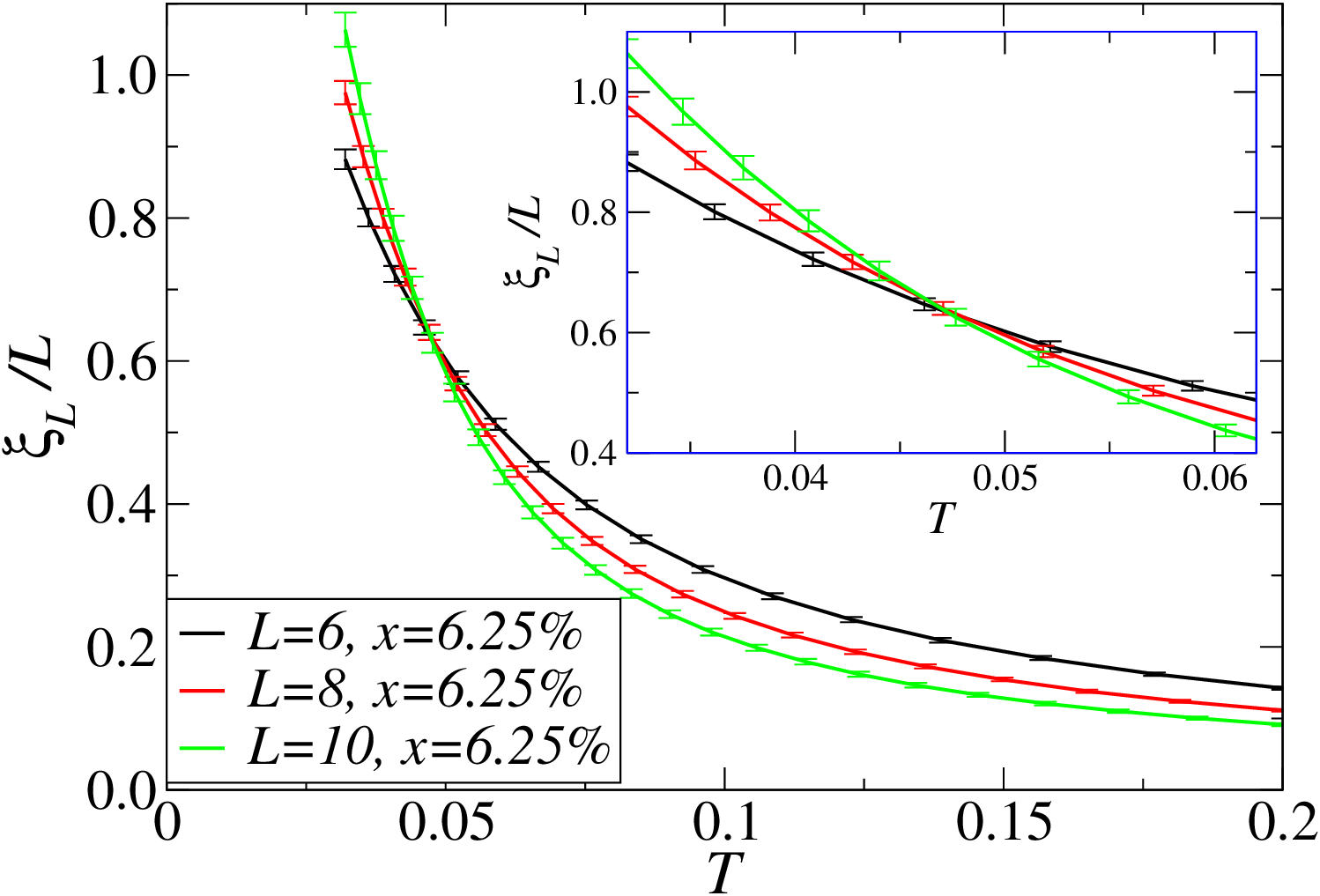}
\includegraphics*[height=0.23\textheight,width=0.39\textwidth, viewport=0 0 750 525,clip]{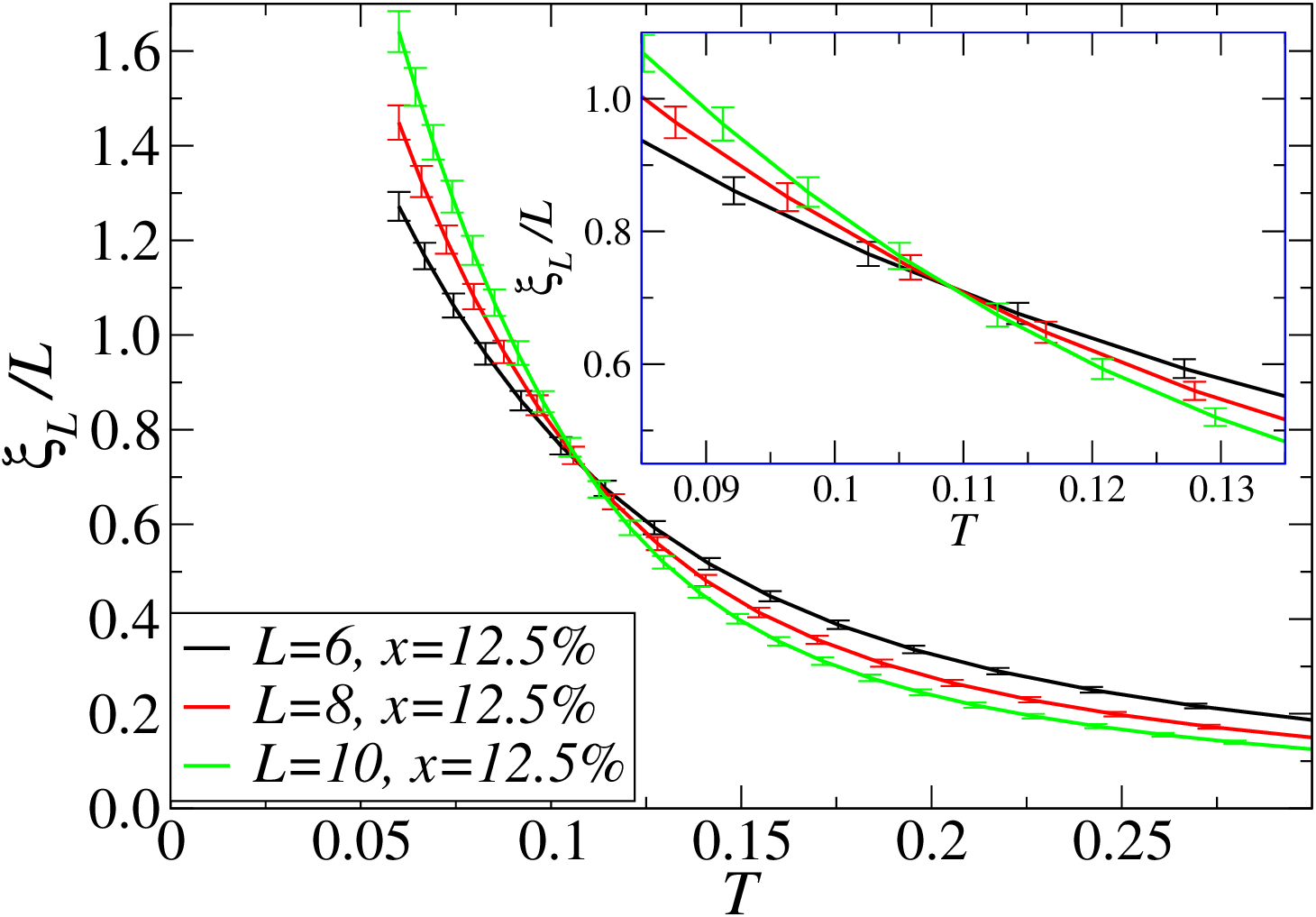}
\caption{Correlation lengths as a function of temperature for $x=6.25\%$(top) and $x=12.5\%$(bottom), where $\xi_{L}/L$  is dimensionless and $T$
is in K units. The insets present the regions close to the crossing temperatures.}
\label{corr}
\end{figure}

Figure \ref{corr} shows $\xi_{L}/L$ vs $T$ for $x=6.25\%$ and $12.5\%$.
A unique and well defined crossing is observed for both concentrations, 
providing compelling evidence that a thermodynamic SG transition at $T_{\rm sg}>0$ 
occurs in model (\ref{Hamiltonian}). 

\begin{figure}[bth]
\centerline{
\includegraphics*[height=0.16\textheight,width=0.24\textwidth, viewport=0 0 750 525,clip]{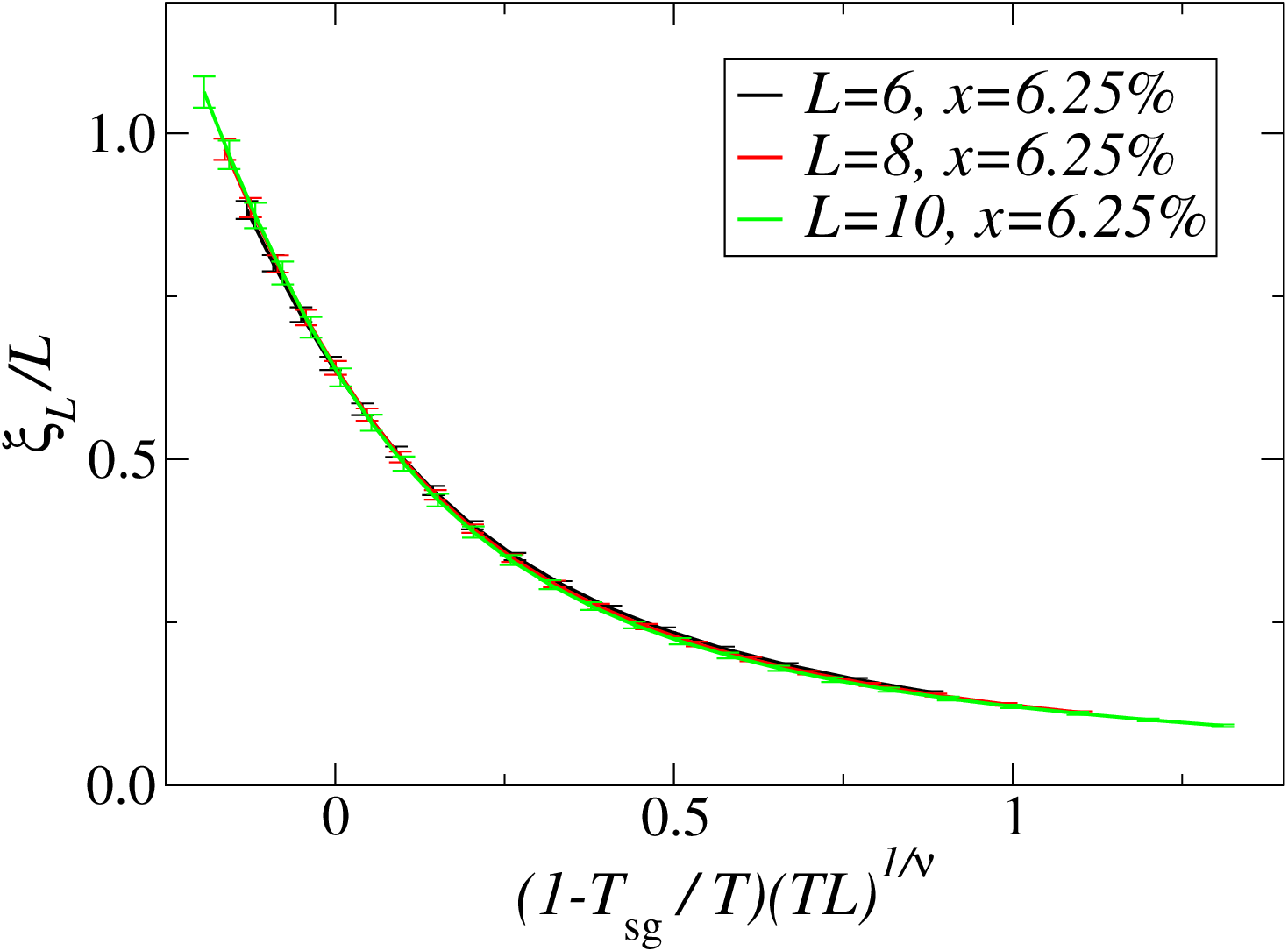}
\includegraphics*[height=0.16\textheight,width=0.24\textwidth, viewport=0 0 750 525,clip]{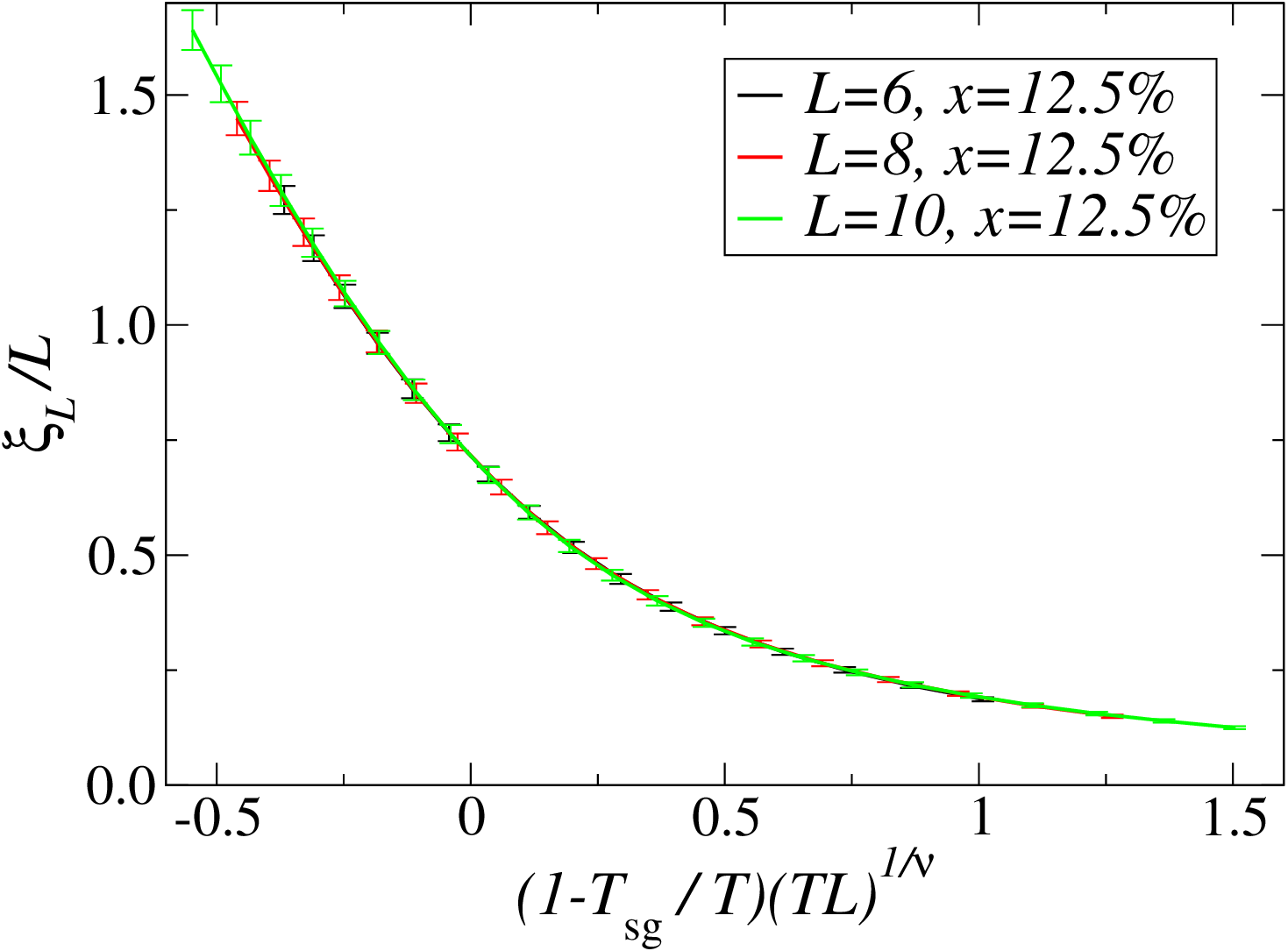}}
\caption{Correlation lengths as a function of $(1-T_{\rm sg}/T)(TL)^{1/\nu}$for $x=6.25\%$(left) and $x=12.5\%$(right), where $\xi_{L}/L$  is 
dimensionless and $T$ is in K units.}
\label{scaled_corr}
\end{figure}

Because of the small systems we need to consider because of
computational constraints, we devised
an extended scaling scheme (ESS)
appropriate for the non-zero mean, 
$[J_{ij}]$, of the  dipolar couplings $J_{ij}$ 
to analyze $\xi_L(L,T)/L$,
$\xi_{L}/L = F[(1-T_{\rm sg}/T)(TL)^{1/\nu}]$ 
\cite{Campbell}.
This ESS is slightly different 
than the one used in Ref.~[\onlinecite{Campbell}]
for the EA model with $[J_{ij}]=0$.
We parameterized the scaling function 
as $F(z)=\sum_{m=0,1,...,4}c_{m}(z-z_{0})^{m}$.  

After estimating $T_{\rm sg}$ from the temperature at which 
$\xi_{L}/L$ cross, the merit function, $\Delta$, 
defined as $\Delta=\sum_{\rm MC \: data}(F(z)L/\xi_{L}-1)^{2}$ 
was minimized to obtain the coefficients $c_{m}$, $z_{0}$ 
and the exponent $1/\nu$. Figure \ref{scaled_corr} shows $\xi_{L}/L$ 
vs the scaling 
parameter $z=(1-T_{\rm sg}/T)(TL)^{1/\nu}$ 
where  $T_{\rm sg} = 0.047$ K, $0.109$ K for $x=6.25\%$ and $12.5\%$, respectively, 
determined from the temperature where the $\xi_{L}/L$ vs $T$  
curves cross in Fig.~\ref{corr}. One finds the scaling exponent 
$1/\nu \approx 0.776$  and $1/\nu \approx 0.782$ for $x=6.25\%$ 
and $12.5\%$, respectively. 
These values are off from $1/\nu \approx 0.37$ for the 3D 
EA Ising model 
with $[J_{ij}]=0$ estimated using an ESS with
 $(1-T_{\rm sg}^{2}/T^{2})(TL)^{1/\nu}$
as scaling parameter \cite{Campbell}. 
One might have expected the critical exponents of the dipolar model (1) 
to be the same as that of the 3D EA model, hence
signaling a common universality class\cite{BM}.
It is likely that the 
simulations of model (\ref{Hamiltonian}) have not yet entered 
the asymptotic finite-size
scaling regime perhaps. 
This is,  in part, because of the proximity 
to the ferromagnetic phase at $x>x_{c}$ 
and the highly spatially anisotropic nature of
the LiHo$_x$Y$_{1-x}$F$_4$ tetragonal unit cell,
which would both introduce 
corrections to scaling not incorporated in $F(z)$.

\begin{figure}[bth]
\centerline{
\includegraphics*[height=0.16\textheight,width=0.24\textwidth, viewport=0 0 750 525,clip]{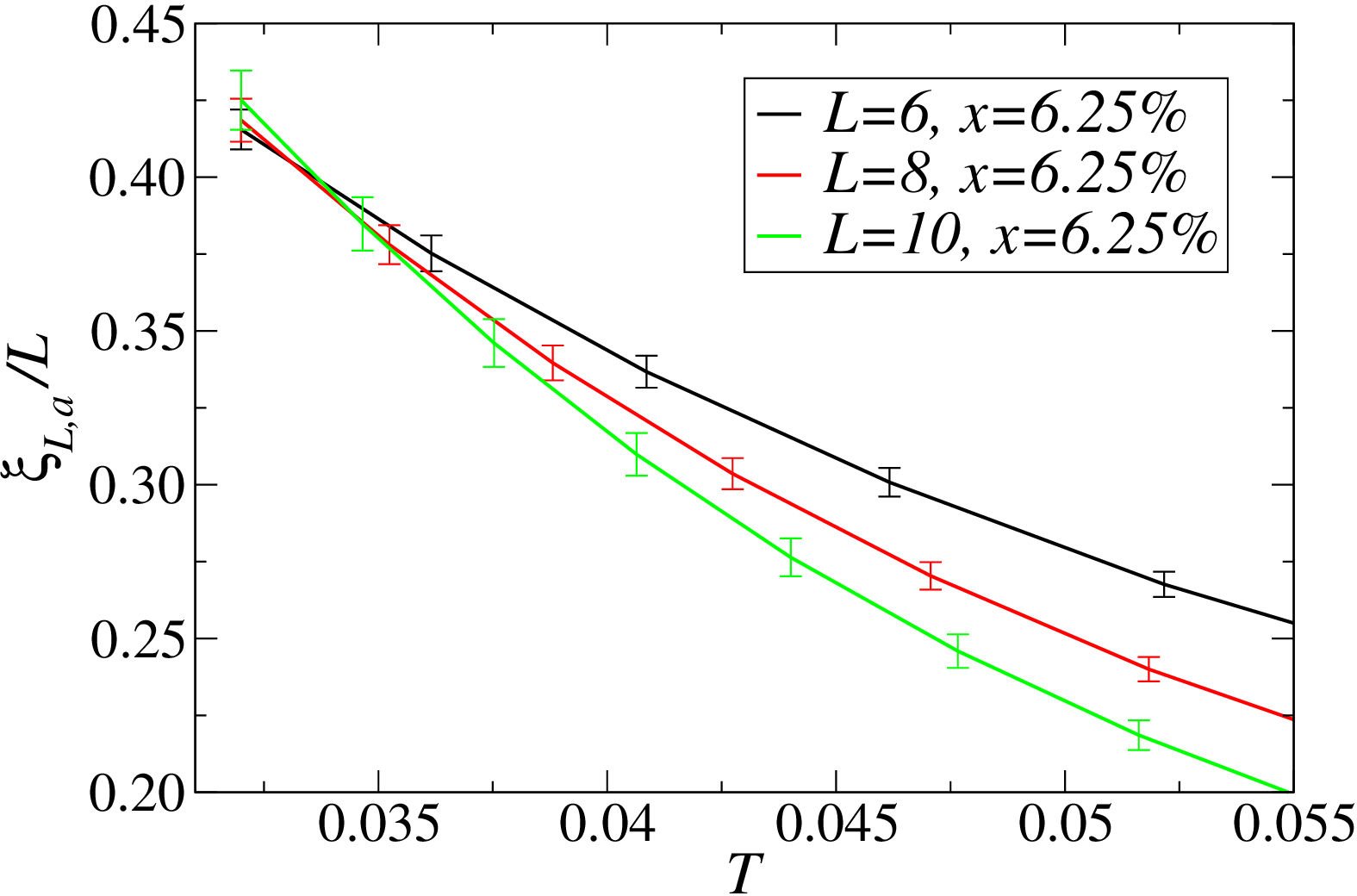}
\includegraphics*[height=0.16\textheight,width=0.24\textwidth, viewport=0 0 750 525,clip]{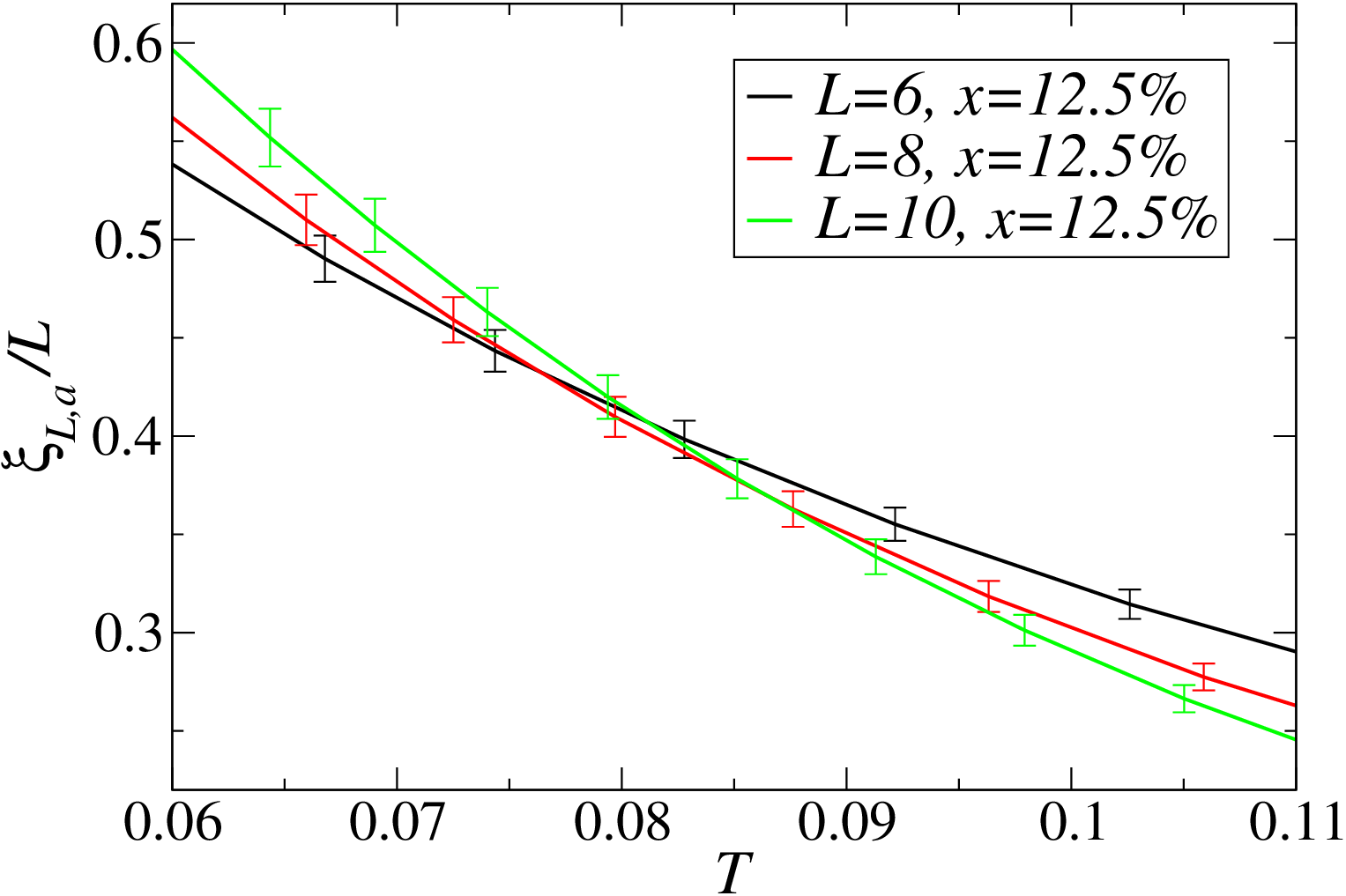}}
\caption{Correlation lengths estimated from $\chi_{\rm sg}(2\pi\hat{x}/(aL))$, $\xi_{L,a}/L$, as a function of temperature for $x=6.25\%$(left) and 
$x=12.5\%$(right), where $\xi_{L,a}/L$  is dimensionless and $T$ is  in K units (only data near the crossing temperatures are shown).}
\label{corrX}
\end{figure}

We now turn to the issue of anisotropic unit cell of
LiHo$_x$Y$_{1-x}$F$_4$.
The Ornstein-Zernike form for $\chi_{\rm sg}$ is at most asymptotically correct. 
The smallest wave vector 
available in our simulation is along the $c$-direction with $\vec{k}=2\pi\hat{z}/(cL)$. 
However,  since LiHo$_{x}$Y$_{1-x}$F$_{4}$ 
is not isotropic, 
it is reasonable to expect that the correlation 
lengths calculated along other directions are not the same as that along the 
$c$-direction. Fig.~\ref{corrX} shows the correlation 
length estimated from $\chi_{\rm sg}(\vec{k}=2\pi\hat{x}/(aL))$.
 We can clearly identify 
crossings at $T=0.034$ K and $T=0.080$ K for $x=6.25\%$ and $x=12.5\%$ respectively, 
which are slightly lower than that from  
$\chi_{\rm sg}(\vec{k}=2\pi\hat{z}/(cL))$. 
We conjecture that, since the couplings among dipoles are stronger 
along the $c$-direction then in the $a$-direction for the
LiHo$_x$Y$_{1-x}$F$_4$ structure, the correlations are enhanced in
the former direction.
This would, for small system sizes, move the $\xi_L/L$ crossings to a relatively
higher temperature than for $\xi_{L,a}/L$.
Here too, important finite-size corrections are likely at play.
However, without access to much larger system sizes and 
without a detailed analysis of the functional form of $\chi_{\rm sg}$, it is 
impossible to explore this anisotropy issue further.

\begin{figure}[tbp]
\centerline{
\includegraphics*[height=0.17\textheight,width=0.25\textwidth, viewport=0 0 750 525,clip]{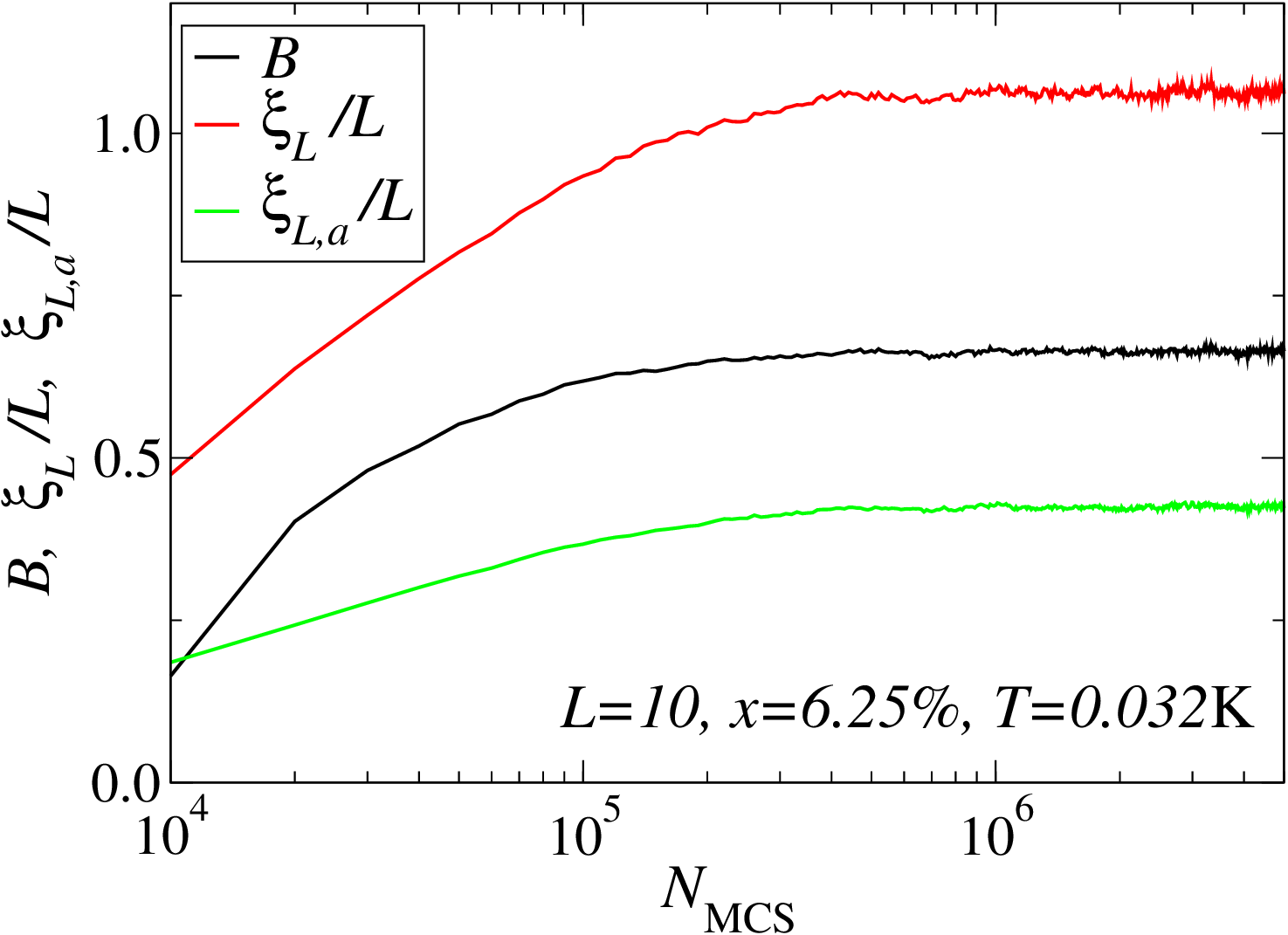}
\includegraphics*[height=0.17\textheight,width=0.25\textwidth, viewport=0 0 750 525,clip]{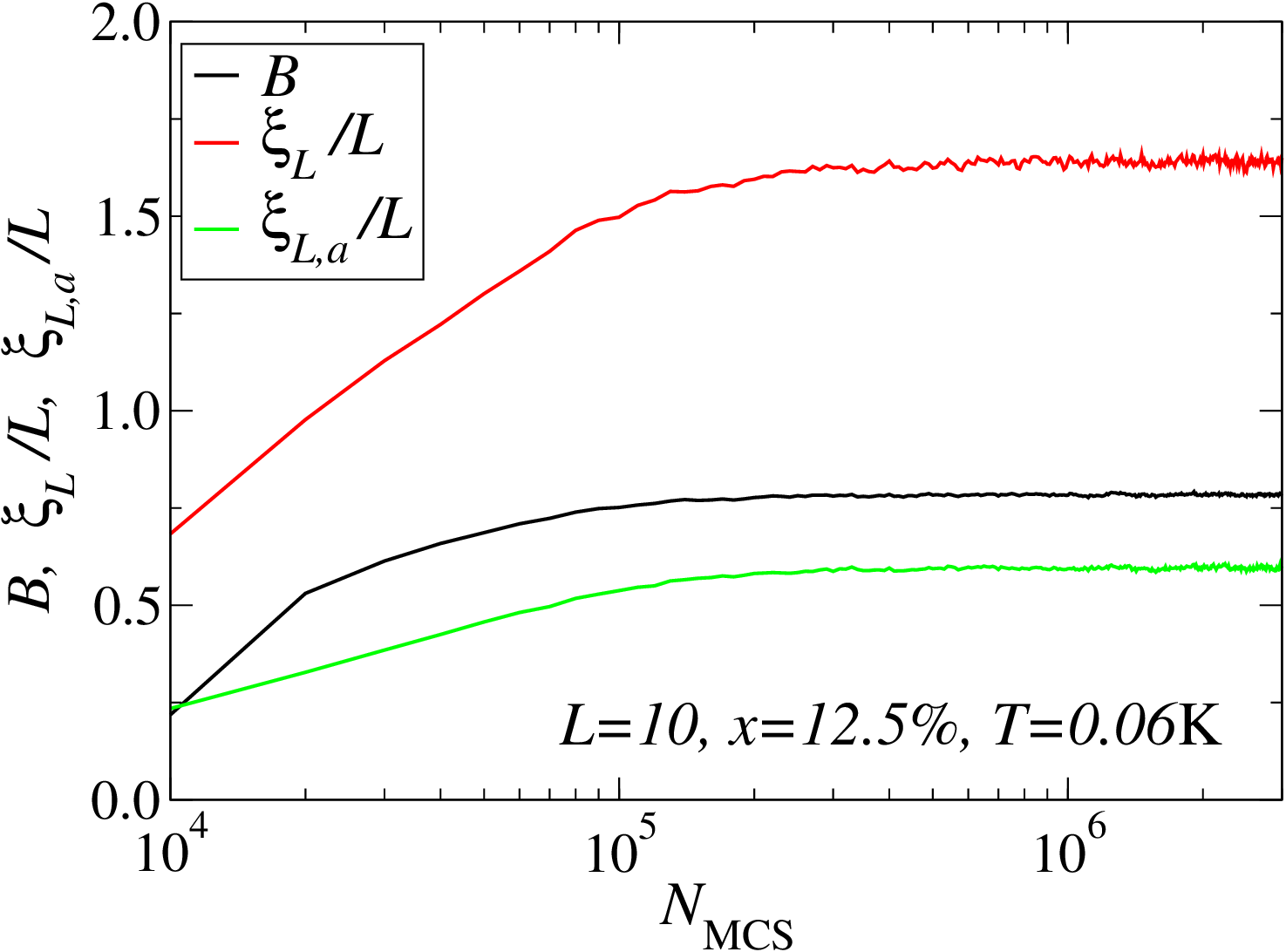}}
\caption{Correlation lengths and Binder ratios as a function of $N_{\rm MCS}$ for $x=6.25\%$(left) and $x=12.5\%$(right) with $L=10$ at the lowest 
temperatures.}
\label{eq}
\end{figure}

The failure of some recent Monte Carlo studies\cite{Biltmo,Snider}
 in identifying a $T_{\rm sg}>0$  transition in model (\ref{Hamiltonian}) is 
mainly because the diluted dipolar system is close to its 
lower critical dimension, as is the 3D EA model,
and because of the sole consideration~\cite{Biltmo} of $B$ as an indicator of
$T_{\rm sg}\ne 0$ as opposed to the more sensitive $\xi_L/L$.
In addition, it is 
difficult to attain equilibrium down to the lowest temperature
 because of the exceedingly slow dynamics. In Fig.~\ref{eq} we show the correlation 
lengths and Binder ratios as a function of $N_{\rm MCS}$
 for the largest system size and the lowest temperature. From the figures, we 
believe that the systems are sufficiently equilibrated for 
extracting reasonably accurate data. 


In summary, we studied a diluted dipolar Ising model of LiHo$_{x}$Y$_{1-x}$F$_{4}$. 
The spin glass (SG)
correlation lengths 
show finite-size crossing as the temperature is lowered as well as scaling behavior,
 providing compelling evidence for a finite-temperature SG
 transition in model (1). It would be desirable to obtain data for
much larger system sizes
 to improve the finite-size scaling analysis. However, aside from 
the very slow spin dynamics upon approaching $T_{\rm sg}$, the computational 
effort scales as $L^6$ due to the long-range nature of the dipolar 
interactions, and simulations of very large system sizes will remain 
prohibitively difficult without a better algorithm. 
Having established that a SG transition occurs in the classical model 
(\ref{Hamiltonian}), and for $x$ as small as $6.25\%$, one may now 
perhaps push further the investigation of the microscopic origin of the  
antiglass state in \lhyf ($x=4.5\%$) \cite{Ghosh,Reich}, assuming that it 
really exists \cite{Quilliam1,Quilliam2}.

We thank P. Henelius, J. Kycia, P. McClarty and P. Stasiak
for useful discussions. This work was funded by
the NSERC, the CRC Program (M.G., tier 1) and SHARCNET. 
We acknowledge SHARCNET and ACE-NET for computational resources.







 
\end{document}